\documentclass[9pt,twocolumn,twoside]{osajnl}

\journal{ol} 

\setboolean{shortarticle}{true}
\usepackage{comment}\allowdisplaybreaks

\title{Selection and cloning of optical patterns with a cold Rydberg atomic gas}
\author[1]{Zeyun Shi}
\author[1,2,3,*]{Guoxiang Huang}

\affil[1]{State Key Laboratory of Precision Spectroscopy, East China Normal University, Shanghai 200062, China}
\affil[2]{NYU-ECNU Joint Institute of Physics, New York University Shanghai, Shanghai 200062, China}
\affil[3]{Collaborative Innovation Center of Extreme Optics, Shangri University, Taiyuan, Shangri 030006, China}
\affil[*]{Corresponding author: gxhuang@phy.ecnu.edu.cn}

\begin{abstract}

We show that optical patterns formed in a cold Rydberg atomic gas working at the condition of electromagnetically induced transparency (EIT) can be
selected by using a weakly modulated control laser field. We also show that the (hexagonal, stripe, square, etc.) patterns prepared in one probe laser field can be cloned onto another one with a high fidelity via a double EIT.

\end{abstract}

\setboolean{displaycopyright}{true}

\begin{document}
\maketitle

{\sl \textbf{Introduction --}}
Spontaneous symmetry breaking and formation of self-organized structures (patterns) in spatially extended systems through dynamical instabilities are important phenomena appearing widely in nature. Well-known examples include convection cells via Rayleigh-B\'{e}nard instability, fluid rolls via Taylor-Couette instability, patterns in nematic liquid crystals via electrohydrodynamic instability, Faraday's crispations via parametric instability, snowflake pattern growth via Mullins-Sekerka instability, structures created in chemical reaction and living systems via Turing instability, and so on~\cite{Cross1993}.

Although the study of pattern formations has a longer history, the exploration in optics is more fundamental since light-matter interaction plays a significant role in many processes. In addition, nonlinear optical devices not only can display phenomena common to most spatially extended systems but also can be used to realize active manipulations on self-organized structures with fast time scales and large frequency bandwidths~\cite{Lugiato2015}.
Due to fundamental interest and potential applications, the research on the generation and control of optical patterns has attracted much attention in recent years~\cite{Lu1996,Martin1996,Mamaev1998,Jensen1998,Wang1998,Kip2000,
Miguez2004,Mau2012,Marsal2017,Bittner2018,Zhang2021}.
Especially, the cloning of localized optical beams (such as solitons) has been considered by using coherent atomic gases~\cite{Vemuri1997,Dey2009,Verma2013,Verma2015,Apolinario2017,Qin2020}, which, however, are valid only for systems with self-focusing Kerr nonlinearity.

In this Letter, we propose a scheme to realize the manipulation and cloning of extended optical patterns in a cold Rydberg atomic gas with a strong, nonlocal self-defocuing Kerr nonlinearity working at the condition of electromagnetically induced transparency (EIT)~\cite{Mohapatra2007}. In the case of three-level ladder-shaped configuration and constant-amplitude control laser field, the system supports only a stable hexagonal pattern, formed through the modulational instability (MI) of a homogeneous plane-wave state. We show that other kinds of optical patterns (stripe, square, etc. which are predicted by the MI but not stable during the  evolution of the system) can be stabilized and selected if the control field is weakly modulated in space. We further show that, by using the nonlocal cross-Kerr nonlinearity, these optical patterns prepared in one probe laser field can be cloned onto another one with high fidelity via a double Rydberg-EIT working in a system with an inverted Y-shaped four-level configuration. The results reported here may have potential applications in all-optical information processing (e.g, optical imaging).

\begin{figure*}[t]
\centering\includegraphics[width=0.95\linewidth]{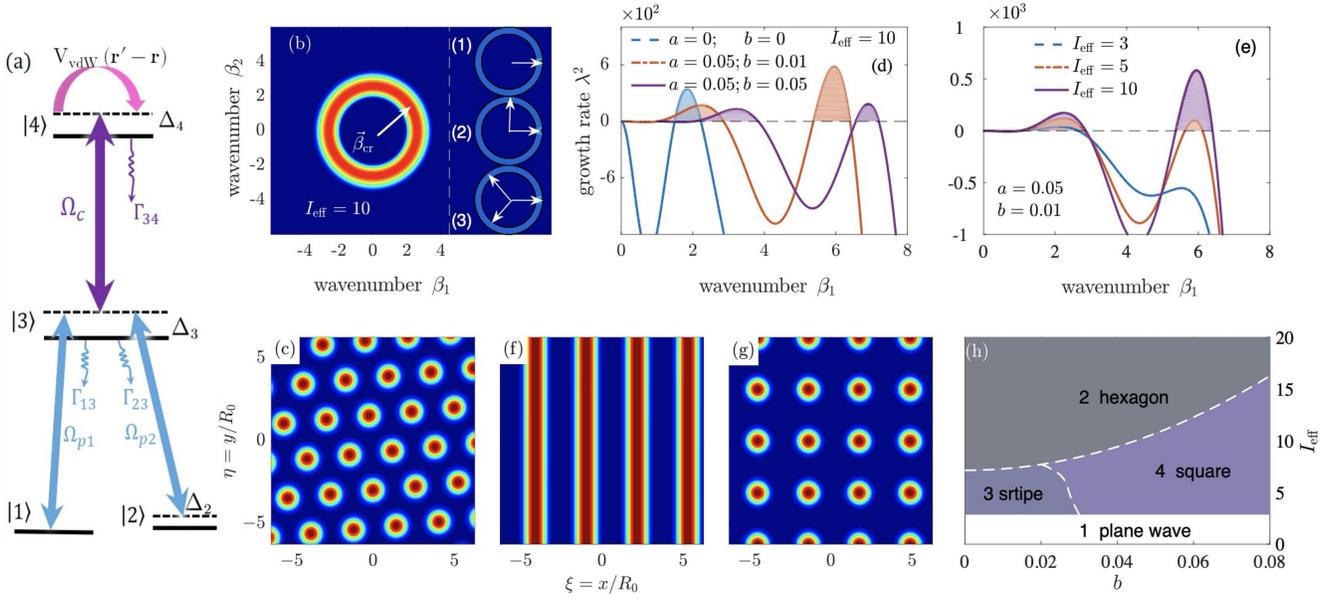}
\caption{\footnotesize
(a)~Inverted Y-shaped excitation scheme supporting the double Rydberg-EIT~(for detail, see text).
(b)~The bright annulus (radius $\beta=\beta_{\rm cr}\approx 2.6$) is the region in 2D momentum space $\vec{\beta}=(\beta_1,\beta_2)$ where MI of the plane-wave state occurs for $I_{\rm eff}=10$ without the modulation of the control field [i.e. $(a,b)=(0,0)$].
Insets: white arrows are wavevectors for possible patterns of hexagon (${\vec \beta}_1+{\vec \beta}_2+{\vec \beta}_3=0$), stripe, and square (${\vec \beta}_1\cdot {\vec \beta}_2=0$).
(c)~Pattern formation: hexagonal lattice structure of the normalized intensity of the probe field 1 ($|u_1|^2$) as a function of  $\xi$ and $\eta$, obtained by  solving Eq.~(\ref{DNLS1}) with the effective probe-field intensity $I_{\rm eff}=10$.
(d)~Growth rate $\lambda^2$  (with $I_{\rm eff}=10$)  as a function of $\beta_1$ for
$(a,b)=(0,0)$ (dashed blue line), $(a,b)=(0.05,0.01)$ (dotted red line), and $(a,b)=(0.05,0.05)$ (solid purple line). The colorful region is the one that MI occurs.
(e)~The same as (d) but with $(a,b)=(0.05,0.01)$ for $I_{\rm eff}=3$ (dashed blue line), 5 (dotted red line), and 10 (solid purple line), respectively.
(f)~Pattern selection (stripe): normalized intensity of the probe field 1 ($|u_1|^2$) as a function of $\xi$ and $\eta$, obtained by solving Eq.~(\ref{DNLS1}) for $(a,b)=(0.05,0.01)$ and $I_{\rm eff}=5$.
(g)~Pattern selection (square lattice): normalized intensity of the probe field 1,
obtained by solving Eq.~(\ref{DNLS1}) for $(a,b)=(0.05,0.05)$ and $I_{\rm eff}=5$.
(h)~Phase diagram for different optical patterns, obtained by fixing $a=0.05$, and varying $I_{\rm eff}$ and  $b$. Regions 1, 2, 3, and 4 are for homogeneous plane-wave, hexagonal, stripe, and square optical patterns, respectively.
}\label{fig1}
\end{figure*}

{{\sl \textbf{Model --}}}
The system we consider is a cold four-level atomic gas with an inverted Y-shaped excitation scheme~[see Fig.~\ref{fig1}(a)].
A weak probe field with angular frequency $\omega_{p1(p2)}$, wavevector ${\bf k}_{p1(p2)}$, and half Rabi frequency $\Omega_{p1(p2)}$ drives the transition from ground state $|1\rangle$~($|2\rangle$) to excited state  $|3\rangle$; a strong control field with angular frequency $\omega_c$, wavevector ${\bf k}_c$, and half Rabi frequency $\Omega_c$ drives the transition from $|3\rangle$  to Rydberg state $|4\rangle$.  $\Delta_{3}$ and $\Delta_{2,4}$ are respectively one- and two-photon detunings; $\Gamma_{13}$, $\Gamma_{23}$, and $\Gamma_{34}$ are spontaneous emission decay rates. The interaction between two Rydberg atoms located respectively at ${\bf r}$ and ${\bf r}'$ is described by van der Waals potential $V_{\mathrm{vdW}}(\mathbf{r}^{\prime}-\mathbf{r}) \equiv \hbar \mathcal{V}(\mathbf{r}^{\prime}-\mathbf{r})$, with $\mathcal{V}({\bf r}^{\prime}-{\bf r})=-C_6/|{\bf r}'-{\bf r}|^6$ [for $^{87}$Rb, $C_6$ is negative, which results in self-defocusing nonlinearity].

The Hamiltonian $\hat{ H}$ of the system can be derived via the electric-dipole and rotating-wave approximations~[for details, see the Sec.~1 of  {\color{blue}Sup.~1}]. The Maxwell-Bloch (MB) equations governing the evolution of the system read
\begin{subequations}\label{MB}
\begin{align}
&\partial\hat{\rho}/\partial t=-(i/\hbar)[{\hat{ H}},\hat{\rho}]-\Gamma\hat{\rho},\label{MBa}\\
&i[\partial/\partial z+(1/c)\partial/\partial t]\Omega_{pj}+c/(2\omega_{pj})\,\nabla_{\perp}^2\Omega_{pj}+\kappa_{j3}
  \rho_{3j}=0,\label{MBb}
\end{align}
\end{subequations}
where $\hat{\rho} ({\bf r},t)$ is density matrix; $\Gamma$ is a $4\times 4$ relaxation matrix describing the spontaneous emission and dephasing; $\nabla_{\perp}^2=\partial^2/\partial x^2+\partial^2/\partial y^2$;  $\kappa_{j3}\equiv\mathcal{N}_a\omega_{p}|{\bf p}_{j3}|^2/(2\varepsilon_0c\hbar)$  ($j=1,\,2$) are coupling constant. Explicit expressions of Eq.~(\ref{MB}) is  presented in the Sec.~1 of {\color{blue}Sup.~1}.

The nonlinear envelope equations controlling the dynamics of the two probe fields can be derived from the MB Eq.~(\ref{MB}) by means of the method of multiple scales beyond mean-field approximation~\cite{Bai2019}. Assuming that $\Omega_{p2}$ is much smaller than $\Omega_{p1}$, we have the dimensionless three-dimensional (3D) {\it nonlocal} nonlinear envelope equations
\begin{subequations}\label{CNNLS}
\begin{align}
&i\frac{\partial u_1}{\partial s}+\tilde{\nabla}_\perp^2 u_1
+\int d^2 \zeta'\, \Re_{11}(\vec{\zeta}-\vec{\zeta}')|u_1(\vec{\zeta}',s)|^2\, u_1=0,\label{DNLS1}\\
&i\frac{\partial u_2}{\partial s}+\tilde{\nabla}_\perp^2 u_2+\int d^2 \zeta'\, \Re_{21}(\vec{\zeta}-\vec{\zeta}')|u_1(\vec{\zeta}',s)|^2\,u_2=0,\label{DNLS2}
\end{align}
\end{subequations}
with $u_j=\Omega_{pj}/U_0$, $s=z/(2L_{\rm diff})$, $\vec{\zeta}=(\xi,\eta)=(x,y)/R_0$, $d^2\zeta'=d\xi'd\eta'$, $\tilde{\nabla}_\bot^2=\partial^2/\partial\xi^2+\partial^2/\partial\eta^2$, and $\Re_{jl}(\vec{\zeta}'-\vec{\zeta})=2L_{\rm diff}R_0^2U_0^2\mathcal{N}'_{jl}[(\vec{\zeta}'-\vec{\zeta})R_0]$. Here $U_0$, $L_{\rm diff}=\omega_pR_0^2/c$, and $R_0$ are typical half Rabi frequency,  diffraction length, and transverse size of the probe field, respectively. For the physical realization of the model described above, the detailed derivation of Eq.~(\ref{CNNLS}), and the expressions of the nonlinear response functions $\Re_{11}$ and $\Re_{21}$, see the Secs.~1 and 3 of \textcolor[rgb]{0,0,1}{Sup. 1}.

{{\sl \textbf{Formation of optical patterns in the probe field 1--}}}
Equation~(\ref{DNLS1}) (which does not involve $u_2$) admits the homogeneous plane-wave solution $u_{\rm pw}(\vec{\zeta},s)=A_0\exp\left[-is A_0^2\int \Re_{11}(\vec{\zeta})d^2 \zeta\right]$, with $A_0$  a real constant. The linear stability analysis can be carried out by assuming that any perturbation to the state is expanded as a linear superposition of many Fourier modes, with the
the growth rate $\lambda$ given by
\begin{align}\label{MI}
\lambda^2=-{\beta}^2\left[\beta^2-2A_0^2\,\tilde{\Re}_{11}(\beta)\right],
\end{align}
where $\vec{\beta}= (\beta_1,\beta_2)$ (${\beta}=[\beta_1^2+\beta_2^2]^{1/2}$) and $\tilde{\Re}_{11}(\beta)$  are the non-dimensional wavevector and nonlinear response function in 2D momentum space; for detail, see  the Sec.~4 of {\color{blue}Sup.~1}.

Shown in the Fig.~S1 of {\color{blue}Sup.~1} is $\lambda^2$ as a function of $\beta_1$ for different effective probe field intensity $I_{\rm eff} \equiv A_0^2\int\Re_{11}(\vec{\zeta})d^2\zeta$. The corresponding growth rate in the 2D momentum space $(\beta_1,\beta_2)$ for $I_{\rm eff}=10$ is given in Fig.~\ref{fig1}(b). The bright annulus [with radius $\beta_{\rm cr}\,\,({\rm critical\,\, wavenumber})\approx 2.6$] in Fig.~\ref{fig1}(b) is the region where MI of the plane-wave state occurs.
Due to the MI, the plane-wave state is unstable, the system will undergo a spontaneous symmetry breaking and the plane-wave state will be transited into self-organized states. Generally, the modes in the bright annulus of Fig.~\ref{fig1}(b) (the MI region) could be the solution of the Eq.~(\ref{DNLS1}), i.e.
$u_1(\vec{\zeta},s)=\sum_{j=1}^N A_j\,e^{i\vec{\beta}_j\cdot \vec{\zeta}}e^{i\mu s}$,
where $\vec{\beta}_j=(\beta_{1j},\beta_{2j})$ ($|\vec{\beta}_j|\approx \beta_{\rm cr}$) and $A_j$ ($j=1,2,...,N$) are complex amplitude functions of $s$, $\xi$, and $\eta$. Because the system has a rotation symmetry, in principal $N$ can take a very large value. However, usually not every one of these modes can really show up in both numerical simulation and experiment. The reason is that the linear MI analysis can  provide only the possibility for the occurrence of the modes in the MI region, but cannot guarantee these modes are stable in the nonlinear evolution of the system (see  the Sec.~4 of {\color{blue}Sup.~1} for more discussions).

Addressing this question requires us to go beyond the MI analysis given above. We hence perform a numerical simulation and seek the ground-state solution of the system, for which the total energy $E = \int |\tilde{\nabla}_{\perp}u_1(\vec{\zeta},s)|^2 d^2\zeta+\frac{1}{2}\iint \Re_{11}({\vec{\zeta}^\prime-\vec{\zeta}})|u_1(\vec{\zeta},s)|^2   |u_1(\vec{\zeta}^{\prime},s)|^2 d^2\zeta^{\prime}  d^2 \zeta$ is minimal. The imaginary-propagation and the split-step Fourier transform methods are used to solve  Eq.~(\ref{DNLS1}). Shown in Fig.~\ref{fig1}(c) is the result of the normalized probe-field intensity $|u_1|^2$ as a function of $\xi$ and $\eta$. Only stable hexagonal structure is found, for which the system supports three wavevectors  satisfying ${\vec \beta}_1+{\vec \beta}_2+{\vec \beta}_3=0$ [they have angle difference of $120^\circ$ with each other; see the inset (3) of Fig.~\ref{fig1}(b)]. In the simulation we have set  $I_{\rm eff}=10$ and $A_0=2.0$ with Gaussian noise as an initial condition.

{{\sl \textbf{Selection of optical patterns for probe field 1--}}}
We now consider the possibility to stabilize and select the more modes predicted by the MI analysis. This is a topic of pattern control widely explored in recent years~\cite{Lu1996,Martin1996,Mamaev1998,Jensen1998,Wang1998,Kip2000,
Miguez2004,Mau2012,Marsal2017,Bittner2018,Zhang2021}. One technique for this is the use of a weak spatial perturbation to some system parameters~\cite{Wang1998}. Although the weak perturbation may modify the system, but the modification does not bring a complete change on system dynamics. In this way, one can obtain an improved performance and desired patterns with accessible system parameters.

Experimentally, the half Rabi frequency of the control field $\Omega_c$ is a parameter control easy to control. Hence, we assume the control field has a weak, periodic spatial modulation
\begin{align}\label{oc}
\Omega_c=\Omega_{c0}[1+a \cos(\xi)+b\cos(\eta)],
\end{align}
where $\Omega_{c0}$ is a real constant, $a$ and $b$ are respectively two small real constants standing for the strengths of the modulation along the $\xi$- and $\eta$-directions. Such spatial modulation may be realized by using a high-resolution space light modulator.

The influence of the modified control field on the MI is checked numerically, with the result plotted in Fig.~\ref{fig1}(d), where the growth rate $\lambda^2$ is plotted (with $I_{\rm eff}=10$) as a function of the wave number $\beta_1$  ($\beta_2=0$) for the modulation parameters of the control field $(a,b)=(0,0)$ (dashed blue line), $(a,b)=(0.05,0.01)$ (dotted red line), and $(a,b)=(0.05,0.05)$ (solid purple line). The colorful region is the one that MI occurs. We see that, comparing Fig.~S1 (for $a=b=0$), MI region is modified due to the influence of the change of the control field. As the increases of the modulation, the critical
wavenumber has a small increase;  additionally, a new MI region (the one on the right hand side) appears.
For non-zero $a$ and $b$, the MI depends also on the value of $I_{\rm eff}$, as shown in Fig.~\ref{fig1}(e) for $(a,b)=(0.05,0.01)$.

The introduction of the weak spatial modulation results in an obvious change of the stability of possible optical patterns. To verify this, a numerical simulation on the dynamical evolution of the system is carried out by solving Eq.~(\ref{DNLS1}) by varying $(a,b, I_{\rm eff})$ from $(0,0,10)$  to $(0.05,0.01,5)$, with other parameters the same as before. Fig.~\ref{fig1}(f) shows the result of $|u_1|^2$ as a function of $\xi$ and $\eta$, obtained by solving Eq.~(\ref{DNLS1}) for $(a,b)=(0.05,0.01)$ and $I_{\rm eff}=5$. In this case
the hexagonal becomes unstable; the system supports a stable stripe pattern, i.e.
it selects only one mode with wavevector $\vec{\beta}_1$ [see the inset (1) in Fig.~\ref{fig1}(b)].
However, when we set $(a,b, I_{\rm eff})=(0.05,0.05,5)$, a stable square structure is observed, as illustrated in Fig.~\ref{fig1}(g), which means that the system selects two orthogonal wavevectors ${\vec \beta}_1$, $ {\vec \beta}_2$  [${\vec \beta}_1\cdot {\vec \beta}_2=0$; see the inset (2) in Fig.~\ref{fig1}(b)].

To find regions where stable patterns of different types are selected by the system, numerical simulations are implemented, which give a phase diagram by fixing $a=0.05$ but varying $b$ and  $I_{\rm eff}$, shown in Fig.~\ref{fig1}(h). In the figure, regions from 1 to 4 are homogeneous plane wave, hexagonal, stripe, and square optical patterns, respectively. One sees that when $b$ is smaller than $a$, the system supports the stripe pattern, but the square pattern can be easily obtained when $a$ and $b$ are about the same magnitude.
When $I_{\rm eff}$ becomes large, the hexagonal pattern becomes preferred one.

The appearance of stripe and square patterns is due to the introduction of the spatially-modulated control field. However, these patterns are self-organized optical structures by the joint action of many factors (including the nonlinear effect contributed by the nonlocal Kerr nonlinearity) in the system, not the result by taking the modulated control field as a strong  ``external'' potential.
The reasons are the following:
(i)~The control field is basically a constant in space because its modulation parameters are very small (i.e. $a\ll 1,\, b\ll 1$).
(ii)~The period of the control field is  $\pi$ both in $\xi$ and $\eta$ directions, but the period  of the (e.g., square) pattern is about $2$ [see Fig.~\ref{fig1}(g)].

The stabilization and selection process of the optical patterns shown above can be understood by the analysis of the total energy of the system, which has been provided in the Sec.~5 of {\color{blue}Sup.~1}. From the result illustrated there, we see that some unstable modes in the absence of the modulation of the control field predicted by the MI analysis can indeed be stabilized. In this sense the patterns in the system can be controlled and selected by the use of the modulated control field. In passing, in addition to the modulated control field, the optical patterns can also be stabilized and selected by using different boundary conditions. For details, see Sec.~6 in {\color{blue}Sup.~1}.

{\sl \textbf{Cloning of hexagonal patterns} --}High-fidelity and controllable optical cloning  are very important for developing novel techniques for optical imaging, lithography, and communications, and so on. In the next, we demonstrate that the optical patterns obtained above can be cloned from one probe field to another one by using the cross-Kerr nonlinearity via the double Rydberg-EIT in the system.

\begin{figure}
\centering\includegraphics[width=1\linewidth]{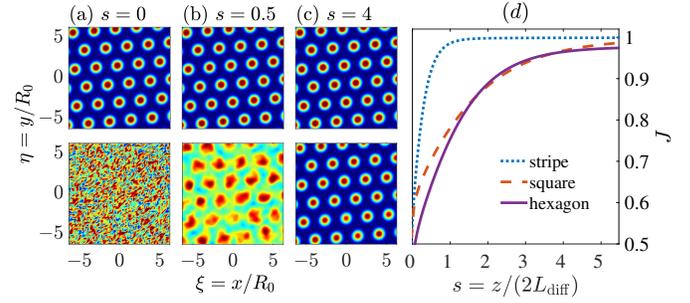}
\caption{\footnotesize Cloning of hexagonal patterns.  Normalized intensities of the probe field 1 (the optical pattern to be cloned; upper row) and the probe field 2 (the optical pattern cloned; lower row) as functions of $\xi$ and $\eta$,  respectively for (a)~$s=0$, (b) $s=0.5$; and (c)~$s=4$.
(d) Waveshape fidelity  $J$  as a function of propagation distance $s$ for different patterns: stripe~(dotted blue line), square~(dashed red line), and hexagon~(solid purple line).
}\label{fig2}
\end{figure}

We first consider the cloning of hexagonal patterns. The process of the cloning is as follows.  At the beginning, a stable hexagonal pattern is obtained through solving  Eq.~(\ref{DNLS1}) for
$I_{\rm eff}=10$ and $a=b=0$, with the result given in the upper row of Fig.~\ref{fig2}(a)-(c),
where $|u_1|^2$ is shown as a function of $\xi$ and $\eta$ when propagating respectively to $s=z/(2L_{\rm diff})=0,\,0.5,$ and 4, with $L_{\rm diff}=0.2$~mm. Then, we use a plane wave adding Gaussian noise as the initial condition [bottom of Fig.~\ref{fig2}(a)] together with a periodic boundary condition  to solve Eq.~(\ref{DNLS2}) to obtain the probe field 2 simultaneously.  We find that the hexagonal pattern of the probe field 1 can be cloned onto the probe field 2 at the propagation distance $s=4$, illustrated in the bottom row of Fig.~\ref{fig2}(a)-(c).  From the figure, we see that the probe field 2 acquires almost the same pattern as the probe field 1. Meanwhile, the  probe field 1 keeps the distribution unchange during propagation.

The quality of the optical pattern cloning realized above can be characterized by the pattern fidelity  $J$ [see the definition given in Sec.~7 in {\color{blue}Sup.~1}]. Shown in Fig.~\ref{fig2}(d) is the
fidelity $J$ is as a function of $s$, illustrated by the solid purple line (for the hexagonal pattern) in the figure.  We see that   $J$ firstly increases and exceeds 0.9 at $s=2$, and reaches  maximal value $J=0.95$ at $s=5$. This means that the hexagonal pattern in the probe field 1 can be well cloned onto the probe field 2 with high fidelity.

\begin{figure}[t]
\centering\includegraphics[width=0.92\linewidth]{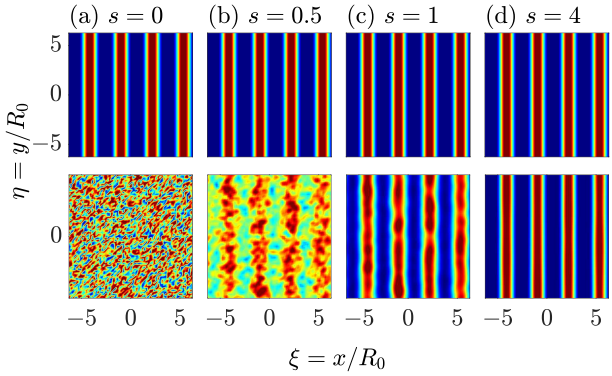}
\caption{\footnotesize Cloning of stripe patterns.  Normalized  intensities of the probe fields 1~(upper row) and 2~(lower row)   as functions of $\xi$ and $\eta$ for propagation distance $s=z/(2L_{\rm diff})=0,\,0.5,\,1,\,4$~($L_{\rm diff}=0.2$~mm), respectively.
}\label{fig3}
\end{figure}

{\sl \textbf{Cloning  of stripe patterns} --}
The next consideration is the cloning of stripe patterns. A stable stripe structure of the probe field 1 is obtained for probe field 1 by solving Eq.~(\ref{DNLS1}) by taking $(a,b)=(0.05,0.01)$ and $I_{\rm eff}=5 $, with the normalized intensity shown in the upper row of Fig.~\ref{fig3}(a)-(d),
respectively for $s=0,\,0.5,\,1$, and $4$. This stripe optical pattern is cloned onto the probe field 2 through solving Eq.~(\ref{DNLS2}). We see that the stripe pattern can also be cloned from the probe field 1 onto the probe field 2 at $s=4$, plotted in the bottom row of Fig.~\ref{fig3}(a)-(d). The fidelity $J$ of the cloning as a function of  $s$ is shown by the dotted blue line of Fig.~\ref{fig2}(d). We see that the cloning fidelity
of the stripe pattern is higher than that of the hexagonal one. It increases with the propagation distance and can reach the maximal value $J=0.96$ at $s=5$.

\begin{figure}[h]
\centering\includegraphics[width=0.92\linewidth]{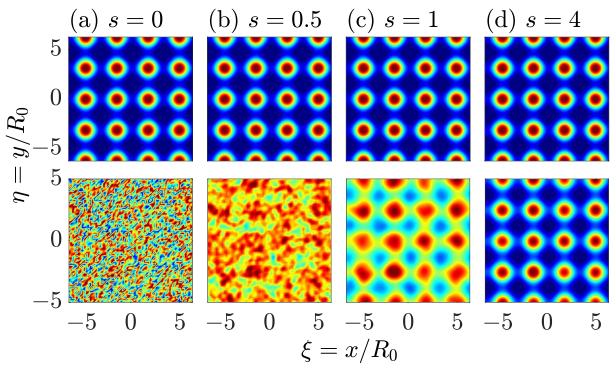}
\caption{\footnotesize Cloning of square patterns.  Normalized  intensities of the probe fields 1 (upper row) and 2 (lower row) as functions of $\xi$ and $\eta$ for propagation distance $s=z/(2L_{\rm diff})=0,\,0.5,\,1,\,4$, respectively.
}\label{fig4}
\end{figure}

{\sl \textbf{Cloning of square patterns} --}
%
Lastly, we consider the cloning of square patterns. Similarly, through the numerical simulation of  Eq.~(\ref{DNLS1}) we obtain the stable square pattern for $(a,b)=(0.05,0.05)$ and $I_{\rm eff}=5$, with the result shown in the upper row of Fig.~\ref{fig4}(a)-(d),
respectively for $s=0,\,0.5,\,1$, and $4$; simultaneously, the square pattern of the probe field 1 is cloned onto the probe field 2 [by solving Eq.~(\ref{DNLS2})] at $s=4$. The cloning fidelity can reach the maximal value $J=0.98$, as shown in Fig.~\ref{fig2}(d).

We have achieved the cloning of the three kinds of optical patterns~(i.e., stripe, square, and hexagonal patterns). Actually, the cloning for other optical patterns can also be realized as long as they are prepared by choosing different modulation ways of the control field. The physical mechanism of the pattern cloning can be understood as follows. After the probe field 1 (i.e., $u_1$, the optical pattern to be cloned) is prepared as a stationary structure, due to the nonlocal cross-Kerr nonlinearity, this pattern will play a role as a nonlocal external potential for the probe field 2 (i.e., $u_2$, the optical pattern cloned) [see Eq.~(\ref{DNLS2})]. Then the probe field 2 is confined and guided stably by the nonlocal external potential contributed by the probe field 1. As a result, the probe field 2 acquires a stable spatial distribution of the probe field 1.

{{\sl \textbf{Conclusion --}}}
We have shown that the optical patterns formed in a cold Rydberg atomic gas with repulsive interaction working at the condition of a double Rydberg-EIT can be controlled and selected by using a spatially-modulated control field. We also show that, by virtue of the cross-Kerr nonlinearity, these optical patterns (stripe, square, and hexagon, etc.) prepared in one probe beam can be cloned onto another one with very high fidelity. The results for the selection and cloning of optical patterns reported here are useful for guiding new experimental findings and promising for practical applications in all-optical information processing and transmission, such as diffraction-free biological and medical imaging.

\begin{backmatter}

\bmsection{Funding}
National Natural Science Foundation of China (11975098).

\bmsection{Disclosures}The authors declare no conflicts of interest.

\bmsection{Data availability} All data needed to evaluate the conclusions in the paper are present in the paper and/or the Supplementary Materials.

\noindent See {\color{blue}Sup. 1} for supporting content.

\end{backmatter}



\end{document}